\documentstyle[aps,epsfig,pre]{revtex}
 \begin{document}
\title{Binder cumulants of an urn model
and Ising model above critical dimension}
\author{Adam  Lipowski$^{1),2)}$ and Michel Droz$^{1)}$}
\address{$^{1)}$ Department of Physics, University of Gen\`eve, CH 1211
Gen\`eve 4, Switzerland\\
$^{2)}$ Department of Physics, A. Mickiewicz 
University, 61-614 Poznan, Poland}
\date{\today}
\maketitle
%%%%%%%%%%%%%%%%%%%%%%%%%%%%%%%%%%%%%%%%%%%%%%%%%%%%%%%%%%%%%%%%%
\begin{abstract}
Solving numerically master equation for a recently introduced urn model, we 
show that the fourth- and sixth-order 
cumulants remain constant along an exactly located line of
critical points.
Obtained values are in very good agreement with values predicted by Br\'ezin 
and Zinn-Justin for the Ising model above the critical dimension.
At the tricritical point cumulants acquire values which also agree with a
suitably extended Br\'ezin and Zinn-Justin approach.
\end{abstract} 
%\begin{multicols}{2}
\tightenlines
%%%%%%%%%%%%%%%%%%%%%%%%%%%%%%%%%%%%%%%%%%%%%%%%%%%%%%%%%%%%%%%%%%%%
%%%%%%%%%% 
 The concept of universality and scale invariance plays a fundamental role in 
 the theory of critical phenomena~\cite{CARDY}.
It is well known that at criticality the system is  characterized by critical
exponents.
Calculation of these exponents for dimension of the system $d$ lower than the 
so-called critical dimension $d_c$ is a highly nontrivial task~\cite{WILSON}.
On the other hand for $d>d_c$ the behaviour of a given system is much 
simpler and critical exponents take mean-field values which are usually simple 
fractional numbers.

However, not everything is clearly understood above the critical dimension.
One of the examples is the Ising model ($d_c=4$) where despite 
intensive research serious discrepancies between analytical~\cite{CHEN} and 
numerical~\cite{BINDER2000} calculations still persist.
Of particular interest is the value of the Binder cumulant at the critical 
point.
Several years ago Br\'ezin and Zinn-Justin (BJ) calculated this 
quantity using field theory methods~\cite{BREZIN} and only recently numerical 
simulations for the $d=5$ model are able to confirm it~\cite{BLOTE}.
Some other properties of the Ising model above critical dimension are still 
poorly explained by existing theories.
For example, the theoretically predicted leading corrections to the susceptibility 
disagree even up the sign with numerical simulations~\cite{BINDER2000}.

In addition to direct simulations of the nearest-neighbour 
Ising model, there are also some other ways to study the critical 
point of Ising model above critical dimension.
For example, Luijten and 
Bl\"ote used the model with $d\leq 3$ but with long-range 
interactions~\cite{LUIJTENBLOTE}.
Using such an approach they confirmed with good accuracy the BJ predictions 
for the Binder cumulant.

In the present paper we propose yet another approach to the problem of 
cumulants above critical dimension.
Namely, we calculate fourth- and sixth-order cumulants at the
critical point of a recently introduced urn model~\cite{LIPDROZ}.
Albeit structureless, this model exhibits a mean-field Ising-type symmetry 
breaking.
Along an exactly located critical line, the obtained values are in a very 
good agreement with values predicted by BJ.
Let us notice that our calculations: (i) are not affected by the inaccuracy 
of the location of the critical point which is a serious problem in the 
case of the Ising model (ii) are based on the numerical solution of the master 
equation which offers a much better accuracy than Monte Carlo simulations.
Moreover, we calculate these cumulants at the tricritical point and show that 
the obtained values are also in agreement with suitably extended calculations 
of BJ.
That  both the Ising model and the (structureless) urn model have the same 
cumulants is a manifestation of strong universality above the upper critical 
dimension: at the critical point not only the lattice structure but also 
the lattice itself becomes irrelevant.
What really  matters is the type of symmetry which is broken and since in both 
cases it is the same $Z_2$ symmetry, the equality of cumulants follows.

Our urn model was motivated by recent experiments on the spatial
separation of shaken sand~\cite{SCHLICH}.
In the present paper we are not concerned with the relation with 
granular matter and a more detailed justification of rules of the urn model 
is omitted~\cite{LIPDROZ}.
The model is defined as follows:
$N$ particles are distributed between two urns A and B and the number of
particles in each urn is denoted as $M$ and $N-M$, respectively.
Particles in a given urn (say A) are subject to thermal fluctuations and the 
temperature $T$ of the urn depends on the number of particles in it as:
\begin{equation}
T(x)=T_0+\Delta(1-x),
\label{temp}
\end{equation}
where $x$ is a fraction of a total number of particles in a 
given urn and $T_0$ and $\Delta$ are positive constants.
(For urn A and B, $x=M/N$ and $(N-M)/N$, respectively.)
Next, we define dynamics of the model~\cite{LIPDROZ}:\\
(i) One of the $N$ particles is selected randomly.\\
(ii) With probability ${\rm exp}[\frac{-1}{T(x)}]$ the selected 
particle changes urns, where $x$ is the fraction of particles in the urn of 
a selected particle.\\

To measure the difference in the occupancy of the urns we define
\begin{equation}
\epsilon=\frac{2M-N}{2N}=\frac{M}{N}-\frac{1}{2}.
\label{order} 
\end{equation}
In the steady state the flux of particles changing their positions
from A to B equals to the flux from B to A.
Since the selected particles are uncorrelated, the above requirement can be 
written as:
\begin{equation}
<M>{\rm exp}[\frac{-1}{T(<M/N>)}]=<N-M>{\rm exp}[\frac{-1}{T(<(N-M)/N>)}],
\label{flux} 
\end{equation}
or equivalently
\begin{equation}
(\frac{1}{2}+<\epsilon>){\rm exp}[\frac{-1}{T(\frac{1}{2}+<\epsilon>)}]=
(\frac{1}{2}-<\epsilon>){\rm exp}[\frac{-1}{T(\frac{1}{2}-<\epsilon>)}].
\label{flux1}  
\end{equation}
Analysis of eq.~(\ref{flux1}) shows~\cite{LIPDROZ} that on the ($\Delta,T_0$) 
phase diagram symmetric ($\epsilon=0$) and asymmetric ($\epsilon\neq 0$)
solutions are separated by the critical line
which is given by the following equation 
\begin{equation}
T_0=\sqrt{\Delta/2} -\Delta/2,\ \ \  0<\Delta<\frac{2}{3}.
\label{critical}
\end{equation}
The critical lines terminates at the tricritical point: 
$\Delta=\frac{2}{3},T_0=\frac{\sqrt 3 -1}{3}$.
Let us notice that a random selection of particles implies basically the 
mean-field nature of this model.
Consequently, at the critical point $\beta=1/2$ and $\gamma\approx 1$ 
(measured from the divergence of the variance of the order parameter), which 
are ordinary mean-field exponents.
However, the calculation of the dynamical exponent $z$ gives 
$z=0.50(1)$~\cite{LIPDROZ} while the mean-field value is 2.
We do not have convincing arguments which would explain such a small  
value of $z$.
Presumably, this fact might be related with a structureless nature 
of our model.

Defining $p(M,t)$ as the probability 
that in a given urn (say A) at the time $t$ there are
$M$ particles, the evolution of the model is described by the following
master equation
\begin{eqnarray}
p(M,t+1)=&&\frac{N-M+1}{N}p(M-1,t)\omega(N-M+1)+
\frac{M+1}{N}p(M+1,t)\omega(M+1)+\nonumber\\
&&p(M,t)\{\frac{M}{N}[1-\omega(M)]+\frac{N-M}{N}[1-\omega(N-M)]\}
 \ {\rm for}\ M=1,2\ldots N-1\nonumber\\
 p(0,t+1)=&&\frac{1}{N}p(1,t)\omega(1)+p(0,t)[1-\omega(N)],\nonumber\\
 p(N,t+1)=&&\frac{1}{N}p(N-1,t)\omega(1)+p(N,t)[1-\omega(N)],
\label{evol1}
\end{eqnarray}
where $\omega(M)={\rm exp}[\frac{-1}{T(M/N)}]$.
Supplementing the above equations with initial conditions one can easily solve 
them numerically.

\begin{figure}
\centerline{\epsfxsize=9cm 
\epsfbox{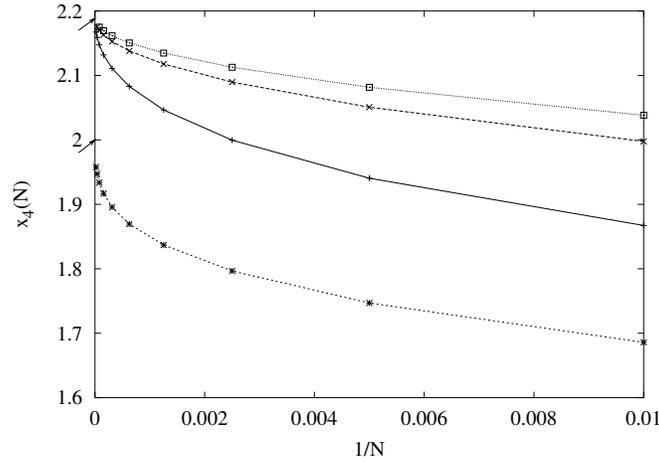}}
%\figspace
\caption{
The fourth-order cumulant $x_4(N)$ as a function of $1/N$ for (from top) 
$\Delta=0.125$, 0.25, 0.5 and $\frac{2}{3}$ (tricritical point).
Arrows indicate the BJ results for the critical and the tricritical 
point.
}
\label{funiversal}
\end{figure}

\begin{figure}
\centerline{\epsfxsize=9cm 
\epsfbox{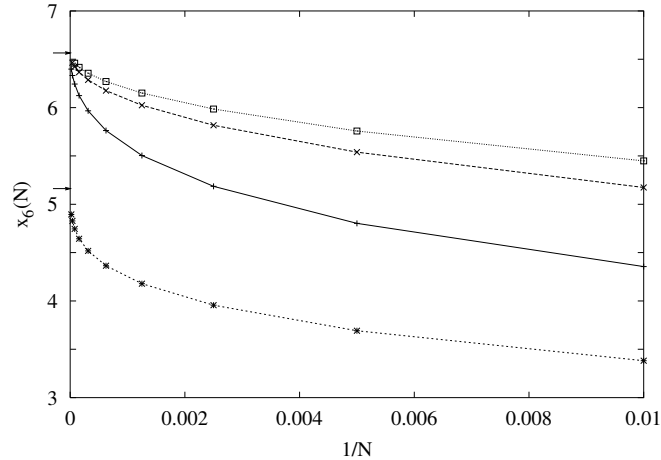}}
%\figspace
\caption{
The same as in Fig.~\ref{funiversal} but for the sixth-order cumulant $x_6(N)$.
}
\label{funiversal1}
\end{figure}

\begin{figure}
\centerline{\epsfxsize=9cm 
\epsfbox{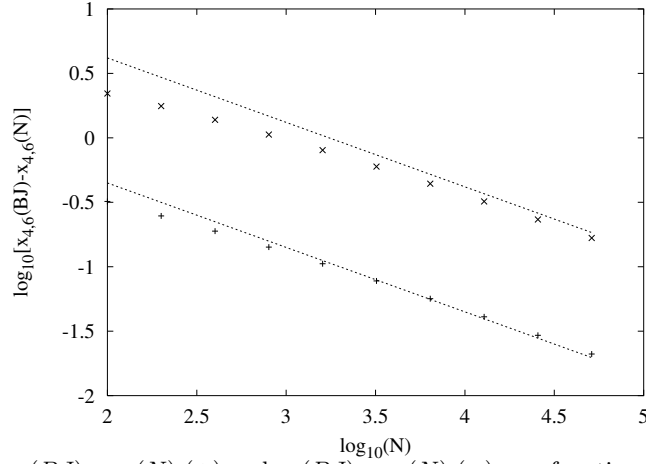}}
%\figspace
\caption{
Logarithmic plot of $x_4(BJ)-x_4(N)$ (+) and $x_6(BJ)-x_6(N)$ 
($\times$) as a function for $N$ for $\Delta=0.5$.
Dotted straight lines have slope 0.5.
}
\label{funiversal05l}
\end{figure}

\begin{figure}
\centerline{\epsfxsize=9cm 
\epsfbox{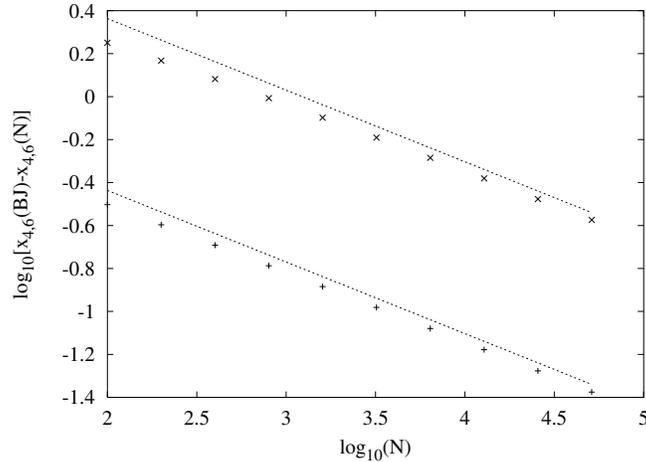}}
%\figspace
\caption{
Logarithmic plot of $x_4(BJ)-x_4(N)$ (+) and $x_6(BJ)-x_6(N)$ 
($\times$) as a function fo $N$ for $\Delta=\frac{2}{3}$ (tricritical point).
Dotted straight lines have slope $\frac{1}{3}$.
}
\label{funiversalttcl}
\end{figure}

Cumulants that we calculate are defined as
\begin{equation}
x_4=\frac{<\epsilon^4>}{<\epsilon^2>^2},\ 
x_6=\frac{<\epsilon^6>}{<\epsilon^2>^3}
\label{cumul}
\end{equation}
where 
\begin{equation}
<\epsilon^n>=\sum_{M=0}^N (\frac{M}{N}-\frac{1}{2})^np(M,\infty)
\label{cumul1}
\end{equation}
and the symbol of infinity indicates that we take the long-time (steady-state) 
solutions of the master equation (\ref{evol1}).
Calculations are made for $\Delta=\frac{1}{8},\frac{1}{4},\frac{1}{2}$ and 
$\frac{2}{3}$ and for each $\Delta$ the value of $T_0$ is calculated from 
eq.~(\ref{critical}).
Thus, the  last point is the tricritical point and the remaining ones are
critical points.
Numerical results are presented in Figs.~\ref{funiversal}-\ref{funiversalttcl}.

Before discussing our results further, let us briefly describe the BJ approach.
To calculate cumulants above the critical dimension they used the 
Ginzburg-Landau-Wilson model.
Then, they calculate the effective action restricting the expansion only to 
the homogeneous contributions (the lowest-mode approximation).
Since at criticality the quadratic (in the order parameter) term vanishes in 
such an expansion and the leading
term is quartic which implies that the probability distribution has the
form $p(x)\sim {\rm e}^{-x^4}$, where $x$ is a rescaled order parameter.
Calculations of moments for such a  distribution are then elementary 
and one obtains
\begin{equation}
x_4=\frac{1}{8\pi^2}[\Gamma(\frac{1}{4})]^4\approx 2.188440...,\,\,\,
x_6=\frac{3}{8\pi^2}[\Gamma(\frac{1}{4})]^{4}\approx 6.565319....
\label{BJ}
\end{equation}
The fact that one can restrict the expansion of the free energy to the lowest 
order term is by no means obvious~\cite{CHEN}.
Such a restriction leads to the correct results but only above critical 
dimension where the model behaves according to the mean-field scenario
with fluctuations playing negligible role.
For $d<d_c$ additional terms in the expansion are also important 
and cumulants take different value.
Numerical confirmation of the above results requires extensive
Monte Carlo simulations, and a satisfactory confirmation was obtained only 
for $x_4$~\cite{BLOTE,BLOTE1}.

Omitting detailed field theory analysis, we can extend the BJ approach to 
the tricritical point.
At such a point also the quartic term vanishes which makes the sixth-order term 
the leading one and the probability
distribution gets the form $p(x)\sim {\rm e}^{-x^6}$.
Simple calculations for such a distribution yield
\begin{equation}
x_4=\frac{\Gamma(\frac{5}{6})\Gamma(\frac{1}{6})}{\Gamma(\frac{1}{2})^2}=2,\,\,\,
x_6=\frac{\Gamma(\frac{1}{6})^3}{6\Gamma(\frac{1}{2})^3}\approx 5.162113....
\label{BJ1}
\end{equation}
%%%%%%%%%%%%%%%%%%%%%%%%%%%%%%%%%%%%%%%%%%%
%%%%%%%%%%%%%%%%%%%%%%%%%%%%%%%%%%%%%%
%%%%%%%%%%%%%%%%%%%%%%%%%%%%%%%%%%%%%
%%%%%%%%%%%%%%%%%%%%%%%%%%%%%%%%%%%%%%
%%%%%%%%%%%%%%%%%%%%%%%%%%%%%%%%%%%%%
The BJ results (\ref{BJ})-(\ref{BJ1}) are indicated by small arrows in 
Figs.~\ref{funiversal}-\ref{funiversal1}.
Even without any extrapolation one can see, especially for critical points, a
good agreement with our results.
Data in Figs.~\ref{funiversal}-\ref{funiversal1} shows strong finite-size
corrections.
To have a better estimations of asymptotic values in the limit 
$N \rightarrow\infty$ we assume finite size corrections of the form
\begin{equation}
x_{4,6}(N)=x_{4,6}(\infty)+AN^{-\omega}.
\label{corr}
\end{equation}
The least-square fitting of our finite-$N$ data to eq.~(\ref{corr}) gives
$x_{4,6}(\infty)$ which agree with BJ values (\ref{BJ})-(\ref{BJ1}) within 
the accuracy better than 0.1\%.
A better estimation of the correction exponent $\omega$ is obtained assuming 
that $x_{4,6}(\infty)$ are given by the BJ values.
The exponent $\omega$ equals then the slope of the date in the logarithmic 
scale as presented in Figs.~\ref{funiversal05l}-\ref{funiversalttcl}.
Our data shows that for the critical(tricritical) point 
$\omega=\frac{1}{2}(\frac{1}{3})$.

Let us notice that leading finite-size corrections to the Binder cumulant in 
the $d=5$ Ising model at the critical point are also of the form $N^{-0.5}$ 
(with $N$ being the linear system size)~\cite{LUIJTENBLOTE}.
Moreover, for the tricritical point but $d<d_c$ 
the probability distribution is known to exhibit a three-peak
structure~\cite{NIELABA}, which is different than the single-peak form
$p(x)\sim {\rm e}^{-x^6}$.
%%%%%%%%%%%%%%%%%%%%%%%%%%%%%%%%%%%%%%
%%%%%%%%%%%%%%%%%%%%%%%%%%%%%%%%%%%%%
%%%%%%%%%%%%%%%%%%%%%%%%%%%%%%%%%%%%%%
%%%%%%%%%%%%%%%%%%%%%%%%%%%%%%%%%%%%%
%%%%%%%%%%%%%%%%%%%%%%%%%%%%%%%%%%%%%%%%%%%%%%%%%%%%%%%%%%%%%%%%%%
In summary, we calculated fourth- and sixth-order cumulants at the critical
and tricritical points in an urn model which undergoes a symmetry 
breaking transition.
Our results confirm that, as predicted by Br\'ezin and Zinn-Justin, the 
critical probability distributions of the rescaled order parameter 
has the form $p(x)\sim {\rm e}^{-x^4}$.
Similarly, for the tricritical point our results suggest that 
$p(x)\sim {\rm e}^{-x^6}$.

Although in our opinion convincing, the results are obtained using numerical
methods.
It would be desirable to have analytical arguments for the generation of
such probability distributions.
It seems that for the presented urn model this might be easier than for the 
Ising-type models.
Let us notice that for the simplest urn model, which was introduced 
by Ehrenfest~\cite{EHREN}, the steady-state probability distribution can be 
calculated exactly in the continuum limit of the master equation and the 
result has the form $p(x)\sim {\rm e}^{-x^2}$, where $x$ is now proportional to the
difference of occupancy $\epsilon$.
In the Ehrenfest model there is no critical point and we expect that a 
distribution of the type ${\rm e}^{-x^2}$ might characterize our model but 
off the critical line (in the symmetric phase).
We hope that when suitably extended, an analytic approach to our model might 
extract critical and tricritical distributions as well.
Such an approach is left as a future problem.
%%%%%%%%%%%%%%%%%%%%%%%%%%%%%%%%%%%%%%%%%%%
%%%%%%%%%%%%%%%%%%%%%%%%%%%%%%%%%%%%%%%%%%%%%%%%%%%%%%%%%%%%%%%%%%%%
%%%%%
\acknowledgements
This work was partially supported by the Swiss National Science Foundation
and the project OFES 00-0578 "COSYC OF SENS".
%%%%%%%%%%%%%%%%%%%%%%%%%%%%%%%%%%%%%%%%%%%%%%%%%%%%%%%%%%%%%%%%%%%%%%%%%%

%%%%%%%%%%%%%%%%%%%%%%%%%%%%%%%%%%%%%%%%%%%%%%%%%%%%%%%%%%%%%%%%%%%%
%%
%\end{multicols}
\end{document}